# Cell jamming and unjamming in development: physical aspects

Ivana Pajic-Lijakovic*, Milan Milivojevic

Faculty of Technology and Metallurgy, Belgrade University, Karnegijeva 4, Belgrade, Serbia

Email: iva@tmf.bg.ac.rs

**Abstract**

Collective cell migration is essential for a wide range of biological processes such as: morphogenesis, wound healing, and cancer spreading. However, it is well known that migrating epithelial collectives frequently undergo jamming, stay trapped some period of time, and then start migration again. Consequently, only a part of epithelial cells actively contributes to the tissue development. In contrast to epithelial cells, migrating mesenchymal collectives successfully avoid the jamming. It has been confirmed that the epithelial unjamming cannot be treated as the epithelial-to-mesenchymal transition. Some other mechanism is responsible for the epithelial jamming/unjamming. Despite extensive research devoted to study the cell jamming/unjamming, we still do not understand the origin of this phenomenon. The origin is connected to physical factors such as: the cell compressive residual stress accumulation and surface characteristics of migrating (unjamming) and resting (jamming) epithelial clusters which depend primarily on the strength of cell-cell adhesion contacts and cell contractility. The main goal of this theoretical consideration is to clarify these cause-consequence relations.

**Glossary of terms**

**Viscoelasticity**: a property of material to show elastic and viscous characteristics when undergo deformation.

**Mechanical stress**: a physical quantity that describes the magnitude of forces per unit area that cause a deformation.

**Normal stress**: can be extensional and compressive. The compressive stress is a type of normal stress caused by shortening of one, two, or three dimensions. This stress can induce a decrease in a system volume. The extensional stress is a type of normal stress caused by extension of one, two, or three dimensions. This stress can induce an increase in a system volume.

**Shear stress**: a type of stress that acts coplanar with the cross section of system.

**Residual stress**: a stress that remain in a system after the original cause of the stress has been removed. It can be normal or shear.

**Stress relaxation:** observed, time-dependent decrease in the stress of a system from the initial value toward the residual stress under constant strain.

**Volumetric strain**: a deformation of system in response to the mechanical stress applied normally.

**Shear strain**: a deformation of system in response to the mechanical stress applied tangentially.

**Strain rate**: a strain change vs. time

**Constitutive models**: stress-strain relationships

**Viscoelastic solids**: a type of viscoelastic behaviour that satisfies following conditions: (1) stress can relax under constant strain in some cases, and (2) strain can relax under constant stress in some cases

**Viscoelastic liquids**: a type of viscoelastic behaviour that satisfies following conditions: (1) stress can relax under constant strain rate in some cases, (2) strain rate can relax under constant stress in some cases, and (3) strain itself cannot relax.

**Tissue surface tension**: a measure of tissue cohesiveness

**Interfacial tension between adjust tissue**: a measure of tissue adhesiveness

**Epithelial-to-mesenchymal transition**: a process by which cells can transit from an epithelial-like phenotype to a mesenchymal-like phenotype.

## 1. Introduction

Accumulating evidence points out that the phenomena of the cell jamming and unjamming represent integral parts of the tissue development (Mongera et al., 208; Blauth et al., 2021; Atia et al., 2018;2021). Cell jamming state transition is the transition from active, contractile (migrating) to passive, non-contractile (resting) state (Pajic-Lijakovic et al., 2019b) and has been recognized within various cellular systems such as: (1) the gastrulation of the developing fruit fly embryo (Bi et al., 2016; Atia et al., 2018; 2021), (2) development of chicken embryo (Spurlin et al., 2019), (3) elongation of the body axis during the zebrafish development (Mongera et al., 2018) and many others. This is a unique characteristic of epithelial cells, while mesenchymal cells avoid the jamming. These cell types also show different characteristics of cell rearrangement. The migrating epithelial collectives have been characterized by an inhomogeneous distribution of cell packing density, velocity, and cell mechanical stress, while mesenchymal cells are prone to establishment more homogeneous structure. In order to deeply understand the origin of these differences, it is necessary to point to the physical factors which govern cell rearrangement.

Cell jamming appears as a consequence of an accumulation of the compressive mechanical stress caused by collective cell migration which induces an increase in cell packing density (Trepat et al., 2009; Atia et al., 2021; Pajic-Lijakovic and Milivojevic, 2021;2022a). Detail description of causes which lead to the accumulation of the compressive stress will be given within this theoretical consideration. An increase in cell packing density results in tissue stiffening, if and only if, cells keep their active contractile state (Pajic-Lijakovic and Milivojevic, 2022a). However, an increase in cell packing density near the cell jamming intensifies contact inhibition of locomotion which results in a weakening of cell-cell adhesion contacts and down-regulation of their propulsion forces (Garcia et al., 2015; Zimmermann et al., 2016). Consequently, non-contractile cells in the resting (jamming) state is softer than contractile (unjamming) ones due to an accumulation of the contractile energy (Kollmannsberger et al., 2011; Schulze et al. 2017; Pajic-Lijakovic and Milivojevic, 2022a). The contractile Madin-Darbvy canine kidney type II (MDCK) cell monolayer has two times larger Young's modulus than non-contractile one (Schulze et al., 2017).

Jamming cell clusters exist some period of time and then undergo unjamming (Pajic-Lijakovic and Milivojevic, 2019a). Mitchel et al. (2020) revealed that cell unjamming transition is distinct from the epithelial-to-mesenchymal transition. This important result points to some other mechanism responsible for the cell unjamming. Despite extensive research to study the cell unjamming, we still do not understand the underlying mechanism. The aim of this review report is to point to the physical factors responsible for the cell unjamming. In order to understand the role of some physical factors, it is necessary to describe the main characteristics of jamming collectives based on the literature.

Since cell packing density vary between cell types and growing conditions, Garcia et al. (2015) introduced other parameters to characterize the cell jamming state such as: (1) cell-cell adhesion, (2) magnitude of cellular forces and persistence time for these forces, and (3) cell shape. It is well known that polarized, contractile, migrating cells have elongated shapes, while non-contractile, resting cells are more rounding (Bi et al., 2016). Intra-cellular forces are primarily a product of cellular contractility, while inter-cellular forces depend on cell rearrangement during collective cell migration. Inter- and intra-cellular forces accompanied by the strength of cell-cell adhesion contacts contribute to the tissue cohesiveness. These parameters are a result of interplay between: (1) cell signalling and gene expression and (2) tissue mechanics (Atia et al., 2021). Although cell and tissue mechanics has been a research focus during the past two decades, we only start to scratch the surface of all the intriguingly complex dependencies between biochemical signalling and gene expression in one hand and mechanical forces and viscoelastic properties on the other.

Consequently, the tissue cohesiveness has been identified as the one of main factors responsible for the local jamming of migrating cell collectives (Steinberg, 1963; Harris, 1976). It depends on the inter-relation between the strength of cell-cell adhesion contacts and cell contractility (Harris, 1976; Pajic-Lijakovic et al., 2023a). Tissue cohesiveness influences the tissue surface tension and viscoelasticity of multicellular systems which have a feedback on cell rearrangement during tissue development (Pajic-Lijakovic et al., 2023a,b,c). While epithelial cells form strong E-cadherin mediated cell-cell adhesion contacts, mesenchymal cells form weak cell-cell adhesion contacts by using other types of cadherin molecules in some cases (Barrga and Mayor, 2019; Devanny et al., 2021). The cell contractility has an opposite effect to the strength of cell-cell adhesion contacts. While the cell contractility enhances the strength of cell-cell adhesion contacts of epithelial cells, it reduces the strength of cell-cell adhesion contacts of mesenchymal cells caused by inter-cellular repulsion (Devanny et al., 2021). Consequently, migrating (unjamming) epithelial cell parts are more cohesive and consequently have a larger surface tension than the resting (jamming) ones (Devanny et al., 2021; Pajic-Lijakovic et al., 2023a; Pajic-Lijakovic and Milivojevic, 2023). As it is mentioned, the migrating epithelium is stiffer and has a distinct rheological behaviour than the resting one (Pajic-Lijakovic and Milivojevic, 2021;2022a). Consequently, epithelial systems can be treated as two-phase systems. One pseudo-phase represents a migrating epithelial sub-population while the other is the jamming epithelial sub-population. The biointerface between the jamming and unjamming epithelial pseudo-phases should be characterized by an interfacial tension (Pajic-Lijakovic et al., 2023c). This parameter has not been measured yet. The main goal of this consideration is to point out the role of this parameter in the process of cell unjamming which hasn't been considered yet and to discuss the way how to measure it. In contrast to the epithelial systems, mesenchymal systems can be treated as mono-phase systems.

Despite extensive research devoted to study cell behaviour under jamming, we still do not understand why mesenchymal cells avoid jamming and what is the physical mechanism which governs unjamming of epithelial parts. The main goal of this consideration is to discuss the phenomenon from the standpoint of biological physics.

**2. Compressive mechanical stress accumulation caused by collective cell migration and cell jamming**

Collective cell migration is an integral part of various biological processes such as: morphogenesis, wound healing, as well as, cancer spreading and has been characterized by the coordination and cooperation (Barriga and Mayor, 2019; Shellard and Mayor, 2019). The coordination is related to the directional cell movement, while the cooperation of migrating collectives is related to the strength of cell-cell adhesion interactions (Shellard and Mayor, 2019). While epithelial cells have high level of coordination and cooperation by migrating in the form of strongly connected cell clusters, mesenchymal cells migrate in the form of weakly connected cell streams and can be characterized by low level of coordination and a medium or high level of cooperation (Clark and Vignjevic, 2015; Shellard and Mayor, 2019). A movement of cell clusters results in their deformation caused by frictional effects with surrounding tissue or extracellular matrix, i.e. extension in the direction of movement and compression in the direction perpendicular to the movement in order to keep their structural integrity, as was shown in **Figure 1** (Pajic-Lijakovic and Milivojevic, 2019a;2020c).

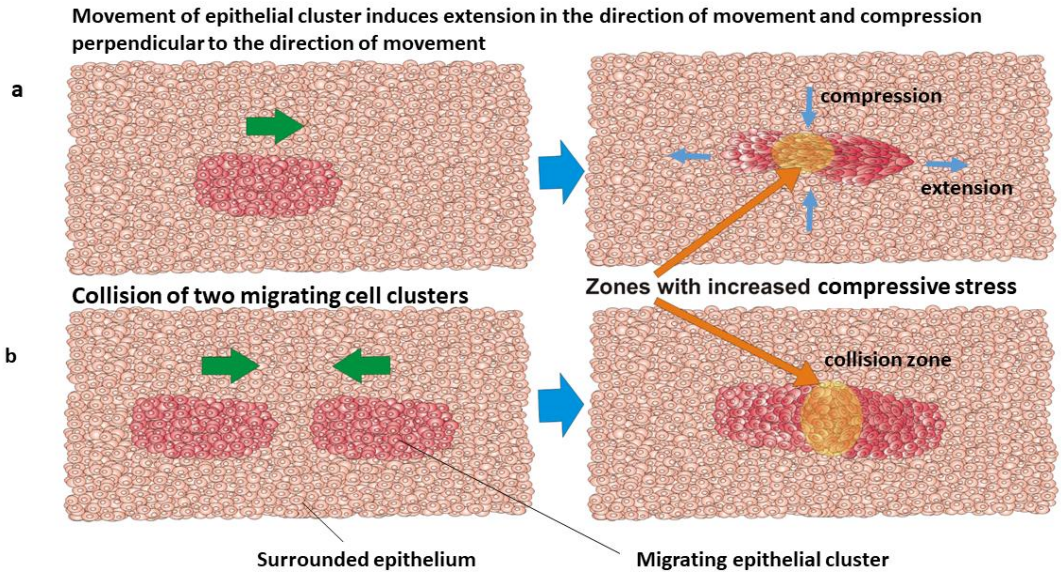

**Figure 1**. Two scenarios which lead to an accumulation of cell compressive residual stress: (a) extension of cell cluster in the direction of movement and compression in the direction perpendicular to movement, and (b) collision of two migrating epithelial clusters caused by uncorrelated motility. Both scenarios represent a consequence of the movement of epithelial cluster through dense surroundings.

Compression of cell clusters can be also induced by collision of migrating clusters caused by uncorrelated motility (Pajic-Lijakovic and Milivojevic, 2019a). The uncorrelated motility is primarily induced by the perturbation of cell signalling. Adjust cells can respond to different signal and/or can behave differently in response to the same signals (Blanchard et al., 2018). Compressive and extensional deformation of a cluster also induces the shear deformation of the cluster parts. This inter-relation between volumetric and shear deformations is a general characteristic of viscoelastic soft matter systems (Pajic-Lijakovic, 2021). The shear deformation is pronounced for the case of: (1) movement of cell streams and (2) effect along the biointerface between adjust tissues (Pajic-Lijakovic and Milivojevic, 2022c). The corresponding strain and the strain rate induce a generation of the mechanical stress (Pajic-Lijakovic and Milivojevic, 2019a). The cell mechanical stress can be normal (extensional or compressive) and shear depending on the corresponding strain. Detail description of the extensional, compressive, and shear stresses accompanied by the corresponding strains and viscoelasticity is given in **Glossary of terms**. Migrating cell collectives show viscoelastic behaviour. It is in accordance with the fact that cell rearrangement caused by collective cell migration induces energy storage and dissipation (Pajic-Lijakovic, 2021). Energy storage and dissipation is a product of the system ability to relax. The stress relaxes toward the equilibrium value which corresponds to the cell residual stress. The residual stress is the stress that remain in a system after the original cause of the stress was removed (**Glossary of terms**). The cell strain changes and residual stress generation occurs in a time scale of hours, while cell stress relaxation occurs in a time scale of minutes (Marmottant et al., 2009; Serra-Picamal et al., 2012; Notbohm et al., 2016; Pajic-Lijakovic and Milivojevic, 2019a). Every displacement increment of cell cluster results in a cell residual stress generation. Cell residual stress can be elastic or dissipative. If cells establish cell-cell adhesion contacts, the residual stress is elastic, while in the opposite case this stress is dissipative. The elastic behaviour of the cell residual stress within migrating epithelial collectives has been confirmed in various model systems such as free expansion of epithelial monolayers and the rearrangement of confluent epithelial monolayers (Serra-Picamal et al., 2012; Notbohm et al., 2016). Serra-Picamal et al. (2012) and Notbohm et al. (2016) revealed that cell residual stress correlates with the corresponding strain which represents a characteristic of the elastic residual stress. The Young's modulus increases with the density and strength of cell-cell adhesion contacts (**Box 1**) (Pajic-Lijakovic and Milivojevic, 2020b). Only elastic cell residual stress can be accumulated within a multicellular system during collective cell migration. Accumulated compressive cell residual stress is the one of the main factors which guides cellular system toward the jamming state (Atia et al., 2021; Pajic-Lijakovic and Milivojevic, 2021). While the migration of epithelial collectives results in a generation of the elastic cell residual stress, the cell residual stress caused by movement of mesenchymal cells is purely dissipative. Elastic residual stress accumulation is a hallmark of viscoelastic solids as was discussed in **Box 1**.

**Box 1. Viscoelasticity of epithelial and mesenchymal migrating collectives**

> Strongly connected migrating epithelial collectives behave as viscoelastic solids. Accordingly, epithelial multicellular systems show following behaviour: (1) mechanical stress can relax under constant strain conditions (Marmottant et al., 2009), (2) strain can relax under constant stress conditions (Mombah et al., 2005; Marmottant et al., 2009), and (3) cell residual stress is elastic (Serra-Picamal et al., 2012; Notbohm et al., 2016). Constitutive model which satisfies all these conditions is the Zener model which has been used for describing the viscoelastic behaviour of migrating epithelial collective with cell packing density lower than or equal to the cell packing density under the confluent state, i.e. $n_c \leq n_{conf}$ (where $n_c$ is the cell packing density and $n_{conf}$ is the cell packing density under the confluent state). It is expressed as (Pajic-Lijakovic and Milivojevic, 2019b; Pajic-Lijakovic, 2021):
>
> $$\tilde{\sigma}_i(r,t,\tau) + \tau_{Ri} \dot{\tilde{\sigma}}_i(r,t,\tau) = G_i \tilde{\varepsilon}_i(r,\tau) + \eta_i \dot{\tilde{\varepsilon}}_i(r,\tau) \qquad (1)$$
>
> where $i \equiv S, V$, $S$ is shear change, $V$ is volumetric change, $r$ is the local space coordinate, $t$ is the time scale of minutes (a short-time scale), $\tau$ is the time scale of hours (a long-time scale), $\tilde{\sigma}_i$ is the shear or normal stress, $\dot{\tilde{\sigma}}_i = \frac{d\tilde{\sigma}_i}{dt}$, $\tilde{\varepsilon}_i$ is the shear or volumetric strain $\tilde{\varepsilon}_S = \frac{1}{2}(\vec{\nabla}\vec{u} + \vec{\nabla}\vec{u}^T)$ and $\tilde{\varepsilon}_V = (\vec{\nabla} \cdot \vec{u})\tilde{I}$, respectively, $\vec{u}(r,\tau)$ is the local cell displacement field caused by CCM, $\vec{v}_c = \frac{d\vec{u}}{d\tau}$ is the cell velocity, $\dot{\tilde{\varepsilon}}_i = \frac{d\tilde{\varepsilon}_i}{d\tau}$ is the strain rate, $G_i$ is the shear or Young's elastic modulus, and $\eta_i$ is the shear or bulk viscosity and $\tau_{Ri}$ is the corresponding stress relaxation time. Stress relaxation under constant strain condition $\tilde{\varepsilon}_{0i}(r,\tau)$ per single short-time relaxation cycle can be expressed starting from the initial condition $\tilde{\sigma}_i(r,0,\tau) = \tilde{\sigma}_{0i}$ as:
>
> $$\tilde{\sigma}_i(r,t,\tau) = \tilde{\sigma}_0 e^{-\frac{t}{\tau_{Ri}}} + \tilde{\sigma}_{Ri}(r,\tau)\left(1 - e^{-\frac{t}{\tau_{Ri}}}\right) \qquad (2)$$
>
> Cell residual stress is elastic and equal to: $\tilde{\sigma}_{Ri} = G_i \tilde{\varepsilon}_{0i}$.
>
> The next step in this theoretical consideration is to emphasize: (1) what is happen with the elastic modulus $G_i$ with weakening of cell-cell adhesion contacts and (2) how to transform the Zener constitutive model to describe the viscoelasticity of migrating mesenchymal collectives. The elastic modulus can be expressed as (Pajic-Lijakovic and Milivojevic, 2020a):
>
> $$G_i \sim \alpha_i \rho_A^{\beta_i} \qquad (3)$$
>
> where $\rho_A$ is the packing density of cell-cell adhesion contacts, $\alpha_i$ is the strength of cell-cell adhesion contacts which depends on the strain (volumetric and shear) and $\beta_i$ is the scaling exponent. A similar relationship between elastic modulus and cohesive contact density within granular systems was expressed by Gaume et al. (2017). Weakening of cell-cell adhesion contacts results in a decrease in the elastic modulus $G_i$. Coordinated movement of cells which are not able to establish cell-cell adhesion contacts satisfies the condition that $G_i \to 0$. In this case, the second term of the eq. 1 can be neglected and the Zener model can be transform to the Maxwell model suitable for describing the rheological behavior of viscoelastic liquids. The Maxwell model has been confirmed experimentally for directional movement of cell streams caused by micropipette aspiration (Guevorkian et al., 2011). Consequently, this constitutive model can be applied for description of the viscoelasticity of collective movement of free cells for the cell packing density lower than or equal to the cell packing density under the confluent state, i.e. $n_c \leq n_{conf}$. The Maxwell model is expressed as:

$$\tilde{\sigma}_i(r,t,\tau) + \tau_{Ri}\dot{\tilde{\sigma}}_i(r,t,\tau) = \eta_i\dot{\tilde{\varepsilon}}_i(r,\tau) \qquad (4)$$

Stress relaxation under constant strain rate condition $\dot{\tilde{\varepsilon}}_{0i}(r,\tau)$ per single short-time relaxation cycle can be expressed starting from the initial condition $\tilde{\sigma}_i(r,0,\tau) = \tilde{\sigma}_{0i}$ as:

$$\tilde{\sigma}_i(r,t,\tau) = \tilde{\sigma}_{0i} e^{-\frac{t}{\tau_{Ri}}} + \tilde{\sigma}_{Ri}(r,\tau)\left(1 - e^{-\frac{t}{\tau_{Ri}}}\right) \qquad (5)$$

where the corresponding cell residual stress is purely dissipative and equal to: $\tilde{\sigma}_{Ri} = \eta_i\dot{\tilde{\varepsilon}}_{0i}$.

Consequently, epithelial-like cells can undergo the jamming state, while the mesenchymal cells are capable to avoiding the jamming. This important result is extracted based on consideration of various model systems such as fusion of two cell aggregates and free expansion of cell monolayers (Nnetu et al., 2013; Heine et al., 2021; Grosser et al., 2021; Pajic-Lijakovic and Milivojevic, 2023). Arrested coalescence during the fusion of two cell aggregates, if exists, is an indicator of cell jamming. Grosser et al. (2021) considered and compared the fusion of: (1) MCF-10A cell aggregates and (2) MDA-MB-436 cell aggregates. While epithelial cells undergo jamming, mesenchymal MDA-MB-436 cells undergo total fusion and consequently avoid jamming. Similar behaviour of epithelial cells also has been shown in 2D cellular systems. While MCF-10A cells frequently undergo jamming during free expansion of cell monolayers (Nnetu et al., 2013; Heine et al., 2021), mesenchymal MDA-MB-231 cells avoid jamming (Heine et al., 2021). The compressive stress of several hundreds of Pa can induce the jamming state transition of 2D epithelial collectives.

### 2.1 The scenario of cell jamming

The cell jamming is driven by the cell compressive residual stress (Atia et al., 2018) based on the scenario shown in **Figure 2**.

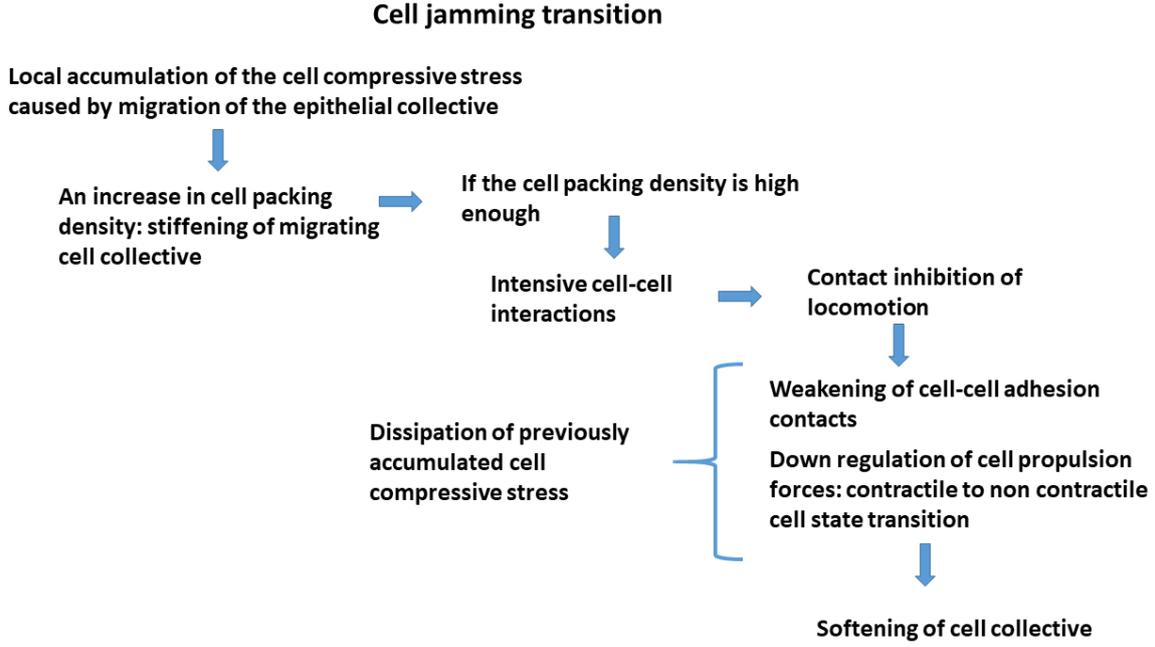

**Figure 2**. Scenario of cell jamming caused by collective cell migration induced by an accumulation of cell compressive residual stress.

The phenomenon can be summarized in several steps: (1) the collective cell migration induces a local increase in the compressive stress, as was shown in **Figure 1** (Pajic-Lijakovic and Milivojevic, 2019a), (2) compressive stress results in an increase in the cell packing density (Trepat et al., 2009), and (3) if an increase in cell packing density is large enough to supress cell movement, cells undergo the jamming state. Cells under the jamming state cannot migrate. They oscillate around their equilibrium positions (Nnetu et al., 2013). These oscillations are damped (Pajic-Lijakovic and Milivojevic, 2019b;2021). The compressive stress caused by collective cell migration, which is capable of inducing the cell jamming transition, is expressed in **Box 2**.

**Box 2**.

> The compressive cell stress, high enough to induce the cell jamming transition, can be expressed by the Fractional constitutive model (Pajic-Lijakovic and Milivojevic, 2019b;2021):
> $$\tilde{\sigma}_i(r,\tau) = \eta_{\gamma i} D^\gamma(\tilde{\varepsilon}_i) \qquad (6)$$
> where $i \equiv S, V$, $S$ is shear change, $V$ is volumetric change, $\eta_{\gamma i}$ is the effective modulus (shear or bulk), $D^\gamma \tilde{\varepsilon}$ are the fractional derivative, and $\gamma$ is the orders of fractional derivatives which satisfy the condition $\gamma < 0.5$. The fractional derivative corresponds to the Caputo's form expressed as:
> $D^\gamma \tilde{\varepsilon} = \frac{1}{\Gamma(1-\gamma)} \frac{d}{d\tau} \int_0^\tau \frac{\tilde{\varepsilon}(r,\tau')}{(\tau-\tau')^\gamma} d\tau'$ (where $\Gamma(1-\gamma)$ is a gamma function and $\tau$ is the time scale of hours) (Podlubny, 1999).

Trapped cells under the jamming state undergo further adaptation to confined micro-environmental conditions (Pajic-Lijakovic and Milivojevic, 2022a). The interactions among trapped cells leads to the contact inhibition of locomotion (Zimmermann et al., 2016). The contact inhibition of locomotion results in the weakening of cell-cell adhesion contacts and reduction of the cell contractility (Iyar et al., 2019. This weakening of cell-cell adhesion contacts results in a dissipation of the cell compressive stress, which is a prerequisite for the cell unjamming. Consequently, the trapped cells under jamming state undergo transformation which accounts for: (1) a decrease in their stiffness, (2) dissipation of the compressive stress and (3) change in the state of viscoelasticity, and surface characteristics which will be discussed in the next section (Devanny et al., 2021; Pajic-Lijakovic et al., 2023a).

## 3. Surface characteristics of epithelial cells and cell unjamming

The jamming and unjamming of cell clusters have been characterized by various values of the tissue surface tension (Devanny et al., 2021). We are interested in macroscopic tissue surface tension which represents a measure of a surface energy of multicellular surfaces in contact with surrounding liquid medium (Pajic-Lijakovic et al., 2023a). The surface energy of multicellular systems accounts for the cumulative effects of the cell-cell adhesion energy and the contractile energy of single cells. Devanny et al. (2021) considered compaction of various contractile and non-contractile MCF-10A cell spheroids and pointed out that the contractility of epithelial cells enhances the strength of E-cadherin mediated cell-cell adhesion contacts, which leads to an establishment of a larger tissue surface tension in comparison with the non-contractile epithelium. In contrast to the epithelial spheroids, mesenchymal spheroids undergo extension rather than compaction which indicates that the surface tension of migrating mesenchymal collectives can be neglected (Devanny et al., 2021; Pajic-Lijakovic and Milivojevic, 2023).

In accordance with the fact that the jamming and unjamming epithelial cell collectives have distinct: (1) viscoelasticity, (2) stiffness, and (3) tissue surface tension, these collectives can be treated as co-existing cell pseudo-phases. The rearrangement of epithelial cells during tissue development is schematically shown schematically in **Figure 3**.

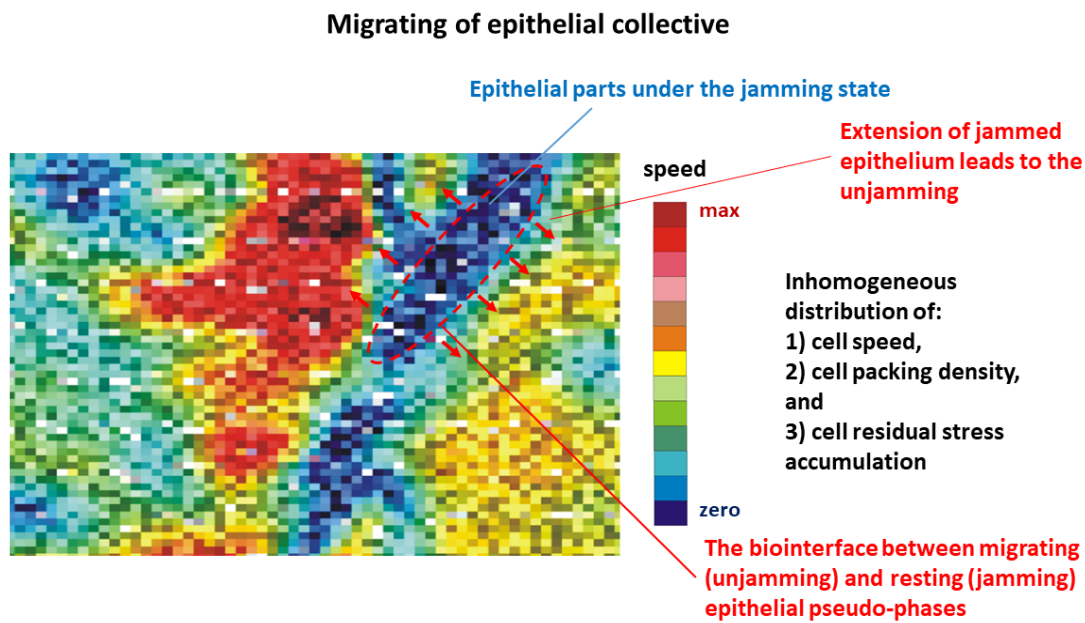

**Figure 3**. Schematic presentation of migrating epithelial collective with jamming cell parts (labelled by blue colour). Inhomogeneous distribution of cell speed is primarily caused by an inhomogeneous distribution of cell packing density. The distribution of cell packing density is a result of inhomogeneous accumulation of cell residual stress.

While majority of epithelial cells migrate, some cell clusters undergo the jamming state. These cell clusters are arrested some period of time and then undergo the unjamming again. These cell jamming-unjamming transitions occur many time during tissue development (Atia et al., 2018). In accordance with the fact that the epithelial jamming and unjamming parts should be treated as the distinct cell pseudo-phases, it is necessary to take into consideration the dynamics along the biointerface between them, which can be characterized by the interfacial tension which represents a product of interactions between migrating and resting cells. These interactions are: (1) biochemical caused by cell signalling and gene expression and (2) mechanical cause by movement of migrating epithelial cells along the biointerface. In further consideration, we will discuss the rearrangement of the pseudo-phases caused by interactions along the biointerface. One pseudo-phase undergoes expansion (wetting) and then compresses the other pseudo-phase.

Isotropic expansion (wetting)/compression (de-wetting) is induced by the work of an interfacial tension described based on the Young-Laplace equation (see eq. 9 in **Box 3**) (Pajic-Lijakovic et al., 2023b). Which pseudo-phase will be extended/compressed depends on the difference between an adhesion energy between the pseudo-phases and cohesion energies of the pseudo-phases themselves. The corresponding quantities are the spreading factors of pseudo-phases. The spreading factor of the pseudo-phase $k$ is equal to $S^k = W_a^{kl} - W_c^k$ (where $W_a^{kl}$ is the adhesion energy between the pseudo-phases $l$ and $k$ equal to $W_a^{kl} = \gamma_l + \gamma_k - \gamma_{lk}$, $\gamma_l$ and $\gamma_k$ are the surface tensions of the pseudo-phases, $\gamma_{lk}$ is the interfacial tension between the pseudo-phases, and $W_c^k$ is the cohesion energy of the pseudo-phase $k$ equal to $W_c^k = 2\gamma_k$). When the spreading factor of the pseudo-phase is larger than zero, this pseudo-phase undergoes extension (Pajic-Lijakovic et al., 2023b). Otherwise, the pseudo-phase undergoes compression. The spreading factor of the jamming (resting) and unjamming (migrating) pseudo-phases is given in **Box 3**.

**Box 3**.

> The spreading factor of the migrating (unjamming) epithelial pseudo-phase is expressed as:
> $$S^{em} = \gamma_r - (\gamma_m + \gamma_{mr}) \qquad (7)$$
> where $\gamma_r$ is the surface tension of resting (jamming) epithelial cells, $\gamma_m$ is the surface tension of migrating (unjamming) epithelial cells, and $\gamma_{mr}$ is the interfacial tension between the pseudo-phases. Both surface tensions accompanied by the interfacial tension between pseudo-phase are space and time dependent (Pajic-Lijakovic et al., 2023a).
> Since the surface tension of resting (non-contractile) epithelial cells is lower than the surface tension of migrating (contractile) epithelial cells ($\gamma_r < \gamma_m$) (Devanny et al., 2021; Pajic-Lijakovic et al., 2023b), the corresponding spreading factor for migrating epithelial cells is lower than zero $S^{em} < 0$. It means that migrating epithelial pseudo-phase undergoes compression. A compression of the migrating epithelial pseudo-phase is directly caused by an extension of the resting epithelial pseudo-phase along the biointerface. Consequently, the spreading factor of the resting (jamming) epithelial pseudo-phase satisfies the condition that $S^{er} > 0$. It is expressed as:
> $$S^{er} = \gamma_m - (\gamma_r + \gamma_{mr}) \qquad (8)$$
> The corresponding isotropic part of the extensional/compressive normal stress can be expressed based on the Young-Laplace equation:
> $$\Delta p_{r \to m} = \pm \gamma_{mr} (\vec{\nabla} \cdot \vec{n}) \qquad (9)$$
> where sign "+" corresponds to an extension, while the sign "-" corresponds to a compression and $\vec{n}$ is the the normal vector of the biointerface.
> Consequently, the extensional residual stress within the resting (jamming) cell pseudo-phase is isotropic and equal to:
> $$\widetilde{\boldsymbol{\sigma}}_{rV}^{er} = +\Delta p_{r \to m} \widetilde{\boldsymbol{I}} \qquad (10)$$

> where $\widetilde{\sigma}_{rV}{}^{er}$ is the cell extensional residual stress and $\widetilde{I}$ is the unit tensor. The residual stress accumulated within a migrating epithelial pseudo-phase includes isotropic contribution caused by interactions along the biointerface and the deviatoric contribution caused by collective cell migration. The total normal stress accumulated within the migrating epithelial pseudo-phase can be expressed as:
>
> $$\widetilde{\sigma}_{rV}{}^{em} = -\Delta p_{r \to m}\widetilde{I} + \widetilde{\sigma}_{erV}{}^{d} \qquad (11)$$
>
> where $\widetilde{\sigma}_{rV}{}^{em}$ is the total cell residual stress within migrating epithelium and $\widetilde{\sigma}_{erV}{}^{d}$ is the deviatoric cell residual stress is the elastic cell residual stress expressed in **Box 1**

The extension of a less cohesive pseudo-phase toward the more cohesive pseudo-phase represents a part of the Marangoni effect (Pajic-Lijakovic and Milivojevic, 2022b). The Marangoni effect has been recognized within various soft matter systems caused by the temperature gradient or the diffusion of some system constituents (Karbalaei et al., 2016).

### 3.1 The scenario of cell unjamming

The cell unjamming is induced primarily by the extension (wetting) of the resting epithelial pseudo-phase toward the migrating epithelial pseudo-phase rather than epithelial-to-mesenchymal transition. During unjamming, cells keep their epithelial phenotype (Mitchel et al., 2020). Consequently, the cell unjamming is influenced by physical parameters such as interplay among the surface tensions of resting and migrating epithelial pseudo-phases, interfacial tension between them and the gradient of interfacial tension. While surface tensions of pseudo-phases depend on homotypic interactions within the migrating and resting epithelial collectives, the interfacial tension depends on heterotypic interactions along the biointerface. These interactions are sensitive to the accumulation of stress (shear and normal) which has a feedback on the remodelling of cell-cell adhesion contacts (Iyer et al., 2019; Pajic-Lijakovic et al., 2023b). Accordingly with fact that the interfacial tension is not constant along the biointerface, we can introduce the gradient of the interfacial tension $\vec{\nabla}\gamma_{mr}$. This gradient is responsible for the cell extension (wetting) along the biointerface which occurs via natural convection. The scenario of the cell unjamming is shown in **Figure 4**.

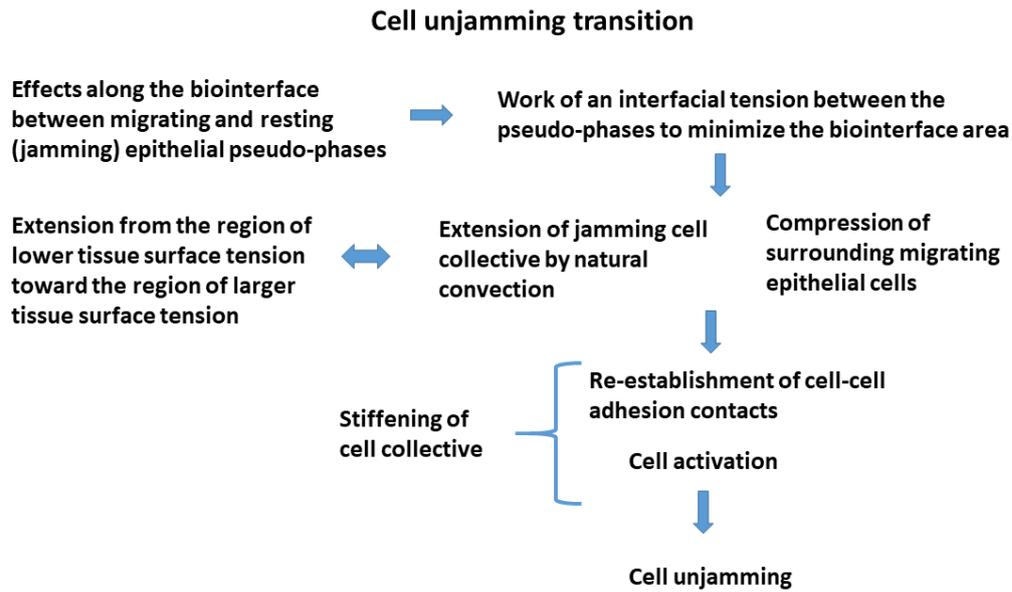

**Figure 4**. Scenario of cell unjamming caused by effects along the biointerface between migrating and resting epithelial pseudo-phases.

The interfacial tension exerts work on reduction the biointerface which results in compression (de-wetting) of the migrating epithelial pseudo-phase and consequently, the extension (wetting) of the resting epithelial pseudo-phase. The extension of non-contractile cells leads to a re-establishment of cell-cell adhesion contacts which induces cell activation and the unjamming again. The remodelling of cell-cell adhesion contacts depends on: a change in: the number of E-cadherin molecules per cell surface and their distribution within the cell surface (Liu et al., 2010; Iyer et al., 2019; Pajic-Lijakovic et al., 2023a). Consequently, the physical parameter such as the interfacial tension between the pseudo-phases is the one of key parameters responsible for the cell unjamming. Despite the importance of this parameter in the cell unjamming, the interfacial tension has not been measured yet.

## 4. The interfacial tension between the epithelial pseudo-phases: measuring techniques

While the tissue surface tension between tissue and surrounding liquid medium has been measured, the interfacial tension between adjust tissues has not been measured yet. Besides the cell unjamming phenomenon considered here, the interfacial tension between adjust tissues influences the ordering of tissues in various compartments during the development (Pajic-Lijakovic et al., 2023b). Several experimental techniques have been used for a measurement of the static tissue surface tension such as: cell aggregate compression between parallel plates (Mombash et al., 2005; Marmottant et al., 2009), cell aggregate micropipette aspiration (Guevorkian et al. (2021), and magnetic force tensiometer (Nagle et al., 2022). However, the tissue surface tension, as well as, the interfacial tension between adjust tissues are time-dependent parameters and it is necessary to measure the dynamic tissue surface tension/interfacial tension (Pajic-Lijakovic et al., 2023a). Inter-relation among various biological processes (which occur at various time scales) such as: (1) the remodelling of cell-cell adhesion contacts, (2) cell signalling, (3) gene expression, and (4) collective cell migration within the region of the multicellular surface/biointerface contribute to the changes of the tissue surface tension (Pajic-Lijakovic et al., 2023a). The static tissue surface tensions measured in the literature depends on the cell type and measured technique. The static tissue surface tension, measured by cell aggregate compression between parallel plates, corresponds to a few $\frac{mN}{m}$ for various cell aggregates (Mombash et al., 2005; Marmottant et al., 2009; Stirbat et al., 2013), while the static tissue surface tension of MCF-10A cell aggregate, measured by magnetic tensiometer corresponds to a few tens of $\frac{mN}{m}$. It is well known that MCF-10A cells form strong cell-cell adhesion contacts which influence the surface tension, but exposure of cell aggregate to magnetic field can additionally enhance the strength of cell-cell adhesion contacts as reported by Jafari et al. (2019). Comprehensive review about the impact of measuring technique on the value of the tissue surface tension doesn't exist.

For the measurement of the (dynamic) interfacial tension between adjust tissues, some non-invasive technique is needed. The magnetic force tensiometer, developed by Nagle et al. (2022), could be also used for the measurement of the interfacial tension by monitoring the temporal change of the biointerface size. Besides the magnetic force, the acoustic force in the form of pseudo-capillary waves can be also applied (Krutyansky et al., 2019; Hobson et al., 2021). The acoustic method for the measurement of an interfacial tension has been applied in various soft matter systems.

## 5. Conclusion and outlook

This theoretical consideration pointed to some physical factors responsible for cell jamming/unjamming caused by collective cell migration such as: compressive cell residual stress, surface tension of jamming and unjamming cell collectives and interfacial tension between them. Only a part of epithelial cells undergoes directional cell movement, while the other part is arrested in the jamming state. Cell clusters in the jamming state represents a physical barrier for the migration of active epithelium and can induce cell swirling motion and collision of migrating cell clusters. In contrast to the epithelial cells, majority of the mesenchymal cells actively migrate and don't undergo the jamming state. The key factor responsible for the difference in cellular behaviour between epithelial and mesenchymal cells is the strength of cell-cell adhesion contacts. While epithelial cells establish strong E-cadherin mediated cell-cell adhesion contacts and migrate in the form of connected cell clusters, mesenchymal cells establish weak adhesion contacts or migrate in the form of cell streams.

Movement of strongly connected cell clusters results in the cell residual stress accumulation, while the movement of free or weakly connected cells are more dissipative. Accumulation of compressive mechanical stress can induce the cell jamming state transition as a consequence of an increase in the cell packing density. This packing density increase intensifies the contact inhibition of locomotion which causes weakening of cell-cell adhesion and the transition from cell active (contractile) to passive (non-contractile) state.

Cell clusters under jamming state show the distinct: (1) viscoelasticity, (2) cluster stiffness, and (3) cohesiveness then the unjamming (migrating) cell clusters and can be treated as a different pseudo-phases. In turn, for the characterisation of the epithelium as a two-phase system, it is necessary to take into consideration: (1) the surface tensions of the migrating and resting cell pseudo-phases, (2) interfacial tension between them, and (3) interfacial tension gradient.

The interfacial tension exerts work on reduction the biointerfacial area. While migrating epithelial cells are compressed, resting epithelial cells (as the less cohesive pseudo-phase) is extended. This extension (wetting) occurs as a natural convection. The interfacial tension gradient is responsible for the cell extension along the biointerface from the region of lower interfacial tension toward the region of larger interfacial tension.

Consequently, this extension of the jamming pseudo-phase induces remodelling of cell-cell adhesion contacts and establishment of cell active (contractile) state again. Despite the fact that an interfacial tension between the jamming and unjamming pseudo-phases play a pivotal role in the cell unjamming, this physical parameter hasn't been measured yet. For deeper understanding the jamming/unjamming of epithelial cells, it is necessary to measure the interfacial tension between the pseudo-phases, as well as the interfacial tension gradient.

**Funding:** This work was supported by the Ministry of Education, Science and Technological Development of the Republic of Serbia (Contract No. 451-03-68/2022-14/200135).

**Declaration of interest:** The authors report there is no conflict of interest.